\documentclass[nonacm]{acmart}
\AtBeginDocument{%
  }

\acmConference[Conference acronym 'XX]{Make sure to enter the correct
  conference title from your rights confirmation emai}{June 03--05,
  2018}{Woodstock, NY}
\acmISBN{978-1-4503-XXXX-X/18/06}

\usepackage{multirow}
\usepackage{xcolor}
\usepackage{eso-pic}
\AddToShipoutPictureFG*{%
  \AtPageLowerLeft{%
    \raisebox{20mm}{%
      \makebox[\paperwidth][c]{%
        \parbox{0.9\paperwidth}{\centering\footnotesize
        This is the author's version of the work. It is posted here for your personal use. Not for redistribution. The definitive Version of Record is published at ACM DL under \url{https://doi.org/10.1145/3816046.3816232}}
        }%
      }%
    }%
  }%

\definecolor{takeawayblue}{HTML}{93BDE5}

\newsavebox{\takeawayboxcontent}

\newenvironment{takeawaybox}
{%
  \par\smallskip
  \setlength{\fboxsep}{8pt}%
  \begin{lrbox}{\takeawayboxcontent}%
  \begin{minipage}{0.92\linewidth}%
}
{%
  \end{minipage}%
  \end{lrbox}%
  \noindent\fcolorbox{takeawayblue}{takeawayblue!20}{\usebox{\takeawayboxcontent}}%
  \par\vspace{1em}
}

\begin{document}

\title{Reaching the last mile? How Design Decisions Shape Medical Device Qualification and Classification of Patient Support Chatbots in the EU}

\renewcommand{\shorttitle}{Reaching the last mile?}

\author{Milo Marsfeldt}
\orcid{0000-0001-5488-919X}
\email{mmsk@create.aau.dk}
\affiliation{%
  \institution{Aalborg University} 
  \country{Denmark}
  \city{Aalborg}
}

\author{Frantisek Kasl}
\orcid{0000-0001-6675-9528}
\email{Frantisek.Kasl@muni.cz}
\affiliation{
    \institution{Masaryk University} 
    \country{Czech Republic}
    \city{Brno}
}

\author{Michal Koščík}
\orcid{0000-0002-9243-6391}
\email{koscik@med.muni.cz}
\affiliation{
    \institution{Masaryk University} 
    \country{Czech Republic}
    \city{Brno}
}

\author{Pavel Loutocky}
\orcid{0000-0002-4965-1467}
\email{loutocky@muni.cz}
\affiliation{
    \institution{Masaryk University} 
    \country{Czech Republic}
    \city{Brno}
}

\author{Vaclav Stupka}
\orcid{0000-0003-1458-1507}
\email{stupka@muni.cz}
\affiliation{
    \institution{Masaryk University} 
    \country{Czech Republic}
    \city{Brno}
}

\author{Robert Mikulik}
\email{mikulik@hotmail.com}
\orcid{0000-0002-7458-5166}
\affiliation{%
  \institution{Health Management Institute}
  \streetaddress{}
  \city{Brno}
  \country{Czech Republic}
  \postcode{}
  }

\author{Hendrik Knoche}
\orcid{0000-0003-3950-8453}
\email{hk@create.aau.dk}
\affiliation{
    \institution{Aalborg University} 
    \country{Denmark}
    \city{Aalborg}
}

\renewcommand{\shortauthors}{Marsfeldt et al.}

\begin{abstract}
Few AI chatbots have transitioned the 'last mile' from research prototypes to clinical practice, partly because mapping conversational design decisions to regulatory requirements is complicated. This paper presents a design-oriented regulatory analysis of a chatbot supporting stroke survivors. Using Design Space Analysis operationalised through the QOC (Questions, Options, Criteria) framework, HCI researchers and legal scholars assess how variations in three conversational functions \textemdash~health Q\&A, recovery monitoring, and motivating medical adherence \textemdash~influence qualification and classification under the European Medical Device Regulation (MDR), General Data Protection Regulation (GDPR), and AI Act (AIA). Our findings show that seemingly minor design decisions \textemdash~such as role, actionability, and proactivity \textemdash~influence regulatory obligations. The paper contributes insights into ten conversational design considerations influencing regulatory interpretation and demonstrates how QOC can support integrating legal considerations into human-centred design workflows. Finally, we suggest compliance-aware pathways to market, while reflecting on whether all chatbots should pursue the 'last mile'.
\end{abstract}



\begin{teaserfigure}
  \centering
  \includegraphics[width=0.8\textwidth]{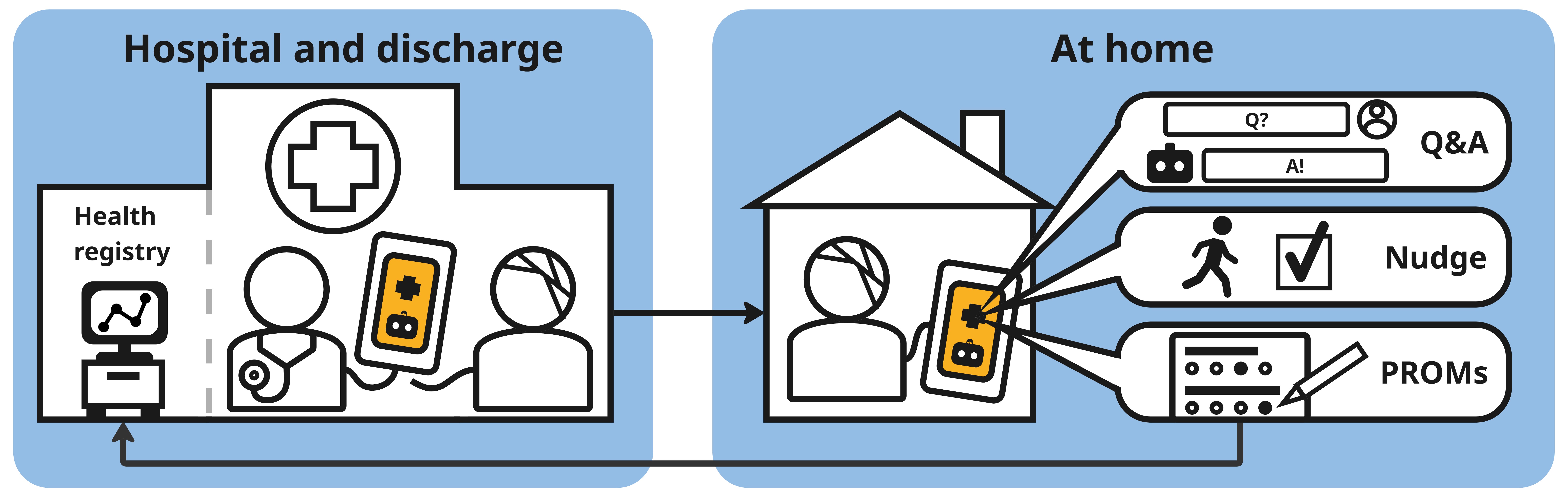}
  \caption{Stroke survivors receive a health chatbot at discharge. At home, they use it to: get answers to health-related questions, receive nudges to follow medical advice, and complete Patient-Reported Outcome Measures (PROM) for hospital monitoring.}\label{Scenario}
  \Description{The diagram shows the scenario of a patient’s care journey from hospital to home. On the left side, there is a hospital building. Inside, a doctor is next to a patient. The doctor gives the patient a smartphone with the chatbot health manager. A computer screen labelled 'Health registry' is also connected to the hospital, showing that patient information is stored and can be used later. On the right side, there is a house representing the patient’s home. Inside is the same patient, still holding the chatbot from which three lines branch out, showing its different features: Q\&A (Questions and Answers): A box with a person asking a question and the app answering. Nudge: A walking figure with a check-mark, representing reminders or encouragement to take action. Patient-Reported Outcome Measures (PROMs): A form with checkboxes and a pen, showing that the patient can fill out questionnaires about their health and recovery. From the PROM, an arrow goes back to the health registry, showing the hospital can monitor the users' answers.
  }
\end{teaserfigure}


\maketitle

\section{Introduction}
Despite years of optimism and considerable investment, most AI systems for healthcare fail to transition from research prototypes to routine clinical use~\cite{Kelly2019, Muehlematter2021} when faced with in-situ clinical workflows, messy real-world data, safety requirements, and regulatory scrutiny~\cite{Zajac2023}. These persistent transition complexities have been dubbed the “last mile” challenge~\cite{Coiera2019}. In a few areas, this is more prevalent than generative conversational AI intended for healthcare~\cite{Cabitza2020, Car2020}. Proposed applications range from pre-screening and diagnostic support~\cite{OteroGonzlez2024, Yenter2021} to acting as patients' at-home health managers~\cite{Milne2020}. Yet, as of September 2024, none of the 1,055 customised generative health chatbots available has received approval from the EU Medical Device Regulation (MDR), the US Food and Drug Administration (FDA), or Australia’s Therapeutic Goods Administration (TGA), despite 59\% of the chatbots impersonating the role of a medical expert \textemdash including professional titles such as 'Doctor', 'Therapist', or 'Nurse' in their name or description~\cite{Chu2025}. This is problematic, as these regulations are not merely procedural checklists, but safeguards intended to protect patients by ensuring systems operate safely, effectively, and reliably. 

Intuitively, one might assume that all chatbots that support health would qualify as medical devices; however, under the law, that is not necessarily the case, as the intended use, context, and marketing all play a role. As a result, small systematic design variations \textemdash~such as role, actionability, and proactivity \textemdash~can determine whether a chatbot falls within regulatory scopes. Consider a health question-and-answer (Q\&A) chatbot, the level of reply personalisation can blur the line between providing general health information and delivering patient-specific medical advice. The former will likely be viewed as a general unregulated well-being application, whereas the latter may be regarded as suggesting a medical treatment, thereby triggering regulations. 

Regulatory-driven studies on medical devices typically adopt either a macro-level perspective \textemdash~explaining how regulations interact~\cite{Aboy2024, Hauglid2023} or questioning whether existing frameworks adequately protect patient safety~\cite{Ebers2024} \textemdash~or a micro-level lens \textemdash~focusing on specific challenges, such as risks associated with generative AI~\cite{Hauglid2023, Chow2025, Azadboni2025}. In contrast, design-driven research emphasises user involvement, usability, and human factors~\cite{Money2011, Privitera2017, Johnson2024}, particularly given the safety risks of device-use errors~\cite{Pelayo2020}. However, this body of work rarely examines how concrete interaction design choices shape regulatory qualification and classification, leaving a disconnect between design practice and regulatory analysis. This gap is particularly problematic because even seemingly small systematic variations in system behaviour may have little effect on obtaining ethical clearance for research prototypes, but can significantly complicate the transition to real-world deployment. Consequently, researchers might be undertaking costly field trials with designs that \textemdash~despite yielding potentially favourable outcomes \textemdash~employ behaviours that might trigger regulatory constraints, preventing the system from reaching the last mile.

This paper presents a multidisciplinary analysis of how conversational design decisions in a healthcare chatbot for stroke survivors (see Figure~\ref{Scenario}) influence regulatory qualification and classification under MDR, GDPR, and the AI Act. Using Design Space Analysis operationalised through the QOC (Questions, Options, Criteria) framework, the study conceptualises three healthcare features \textemdash~1)~health Q\&A, 2)~recovery monitoring, and 3)~motivating medical adherence \textemdash~into ten generalisable design questions. The paper contributes: 1)~insights into how conversational design decisions, such as proactivity and data use, systematically influence regulatory interpretation and compliance obligations. 2)~a QOC-based approach for conceptualising regulatory assessment as a conversational design space within human-centred design workflows. Finally, the paper discusses practical strategies for approaching regulatory compliance and clinical deployment.

\section{Background}
Long before Coiera~\cite{Coiera2019} framed healthcare AI deployment as a “last mile” problem, researchers had argued that successful clinical adoption depends on broader sociotechnical factors beyond model performance alone~\cite{Berg1999}. Coiera highlighted three dimensions: measuring real performance, calibrating models to an ever-changing setting, and adapting to local contexts~\cite{Coiera2019}. Subsequent work has expanded these dimensions, for example, Cabitza et al.~\cite{Cabitza2020} described two sociotechnical chasms affecting deployment: the hiatus of human trust and the hiatus of machine experience. Zajac et al.~\cite{Zajac2023} identified five sociotechnical challenges, including political and legislative factors, and concluded that external regulation is a decisive issue during deployment in clinical environments. Despite this, few studies actively include "legality" as a design dimension when exploring the sociotechnical design space.

\subsection{Designing legally compliant health systems}
Bridging the gap between legal requirements and technical implementation has been a central concern in the literature on designing legally compliant systems. A central challenge identified across this work is that developers often lack legal domain expertise to interpret and implement regulatory requirements, while legal experts lack the technical knowledge to assess system implementations~\cite{dickhaut2022role}. To address this, several approaches have emerged. Pattern-based methods have proven particularly promising: Hoffmann et al.~\cite{hoffmann2012towards} developed software requirement patterns to support the elicitation of legal aspects during requirements engineering, while Dickhaut et al.~\cite{dickhaut2022role} demonstrated how design patterns can provide twofold value by supporting both developers in implementing legal requirements and legal experts in assessing system lawfulness, effectively creating a shared language between disciplines. However, these approaches typically address compliance verification as a post-hoc activity or focus on translating legal requirements into technical specifications, rather than examining how fundamental design decisions at the interaction level systematically shape regulatory outcomes from the outset of development.

\subsection{Case: Stroke survivors}
A stroke occurs when blood flow to the brain is disrupted, potentially causing permanent damage. Resulting in over half of stroke survivors experience long-term physical and cognitive consequences~\cite{Feigin2022, Rouillard2012}. Despite these impairments, economic pressures force healthcare systems to focus primarily on survival, stabilisation, and rapid discharge~\cite{Rochmah2021}, pushing most recovery to the home. Consequently, many stroke survivors and their families feel underprepared and abandoned~\cite{Tholin2014, Martinsen2015}. Simultaneously, healthcare systems lack comprehensive and systematic post-discharge data~\cite{Wissel2013} needed to understand the long-term consequences of in-hospital care~\cite{Bebber2018} and to enable proactive interventions when patients’ health deteriorates or when deterioration is likely due to non-adherence to treatment regimens. Health chatbots offer a scalable way to extend continuity of care beyond discharge~\cite{MacNeill2024}. 

\subsubsection{Health Q\&A}
Because stroke occurs suddenly, survivors and their caregivers face urgent information gaps, from understanding the condition to managing rehabilitation. Yet finding reliable and comprehensible health information online can be difficult~\cite{Fashi2022}. Reflecting this need, 56\% of generative health chatbots include patient-facing Q\&A features~\cite{Chu2025}, functioning as multilingual interactive encyclopedias that efficiently deliver tailored, user-friendly, and empathetic responses~\cite{Rey2024}. Unsurprisingly, both clinicians and patients increasingly consult conversational generative AI systems to seek information and advice~\cite{Shahsavar2023, Ozkan2025} \textemdash~despite these systems not being medical devices. However, generated advice can be misleading, incorrect, or dangerous~\cite{Eichenberger2025, Hu2025}. For example, generative models can produce hallucinations \textemdash~plausible but incorrect statements \textemdash~and may reproduce biases in their training data, raising concerns about harmful discrimination~\cite{Hicks2024, Shachar2023}.

\subsubsection{Recovery monitoring} 
Improve stroke care increasingly requires long-term post-discharge monitoring~\cite{Carod-Artal2009}, combining objective measurements (e.g., blood pressure) with subjective assessments, such as patient-reported outcome measurements (PROMs)~\cite{Burstin2019, Baumhauer2017}. PROMs use validated structured questionnaires to map patients' subjective health experiences into comparable numerical values, without third-party interpretation~\cite{FDA2009}. Enabling outcome evaluation at both individual and population levels~\cite{Hjollund2019}, by comparing patients' continued health~\cite{Black2013} or different treatments~\cite{Patrick2007, Baumhauer2017}. 

Chatbot-facilitated PROM collections show promise, as studies report higher response rates (79–90\%) than web-based PROMs~\cite{Gomaa2023, Chaix2022, GuedeFernndez2024, Kupper2024}, while maintaining answer consistency~\cite{Maharjan2021, Anthony2017, Skovfoged2026}. Participants also found chatbots reduced response fatigue~\cite{Yun2024}, and cognitive demand~\cite{Ponathil2020, Kim2019}, while being more user-friendly, intuitive, and enjoyable~\cite{Celino2020, Soni2022, tePas2020, Kocielnik2021} compared to web-based surveys. Yet, no chatbot administering PROM has yet obtained MDR certification. 

\subsubsection{Health nudges}
Stroke survivors' continued well-being and risk of recurring strokes are directly linked to their adherence to medical prevention, treatment, and rehabilitation regimens~\cite{Kelley2009}. However, adapting new regimens requires modifying pre-stroke habits and routines~\cite{Badawy2020}. Chatbots can support this process by offering digital nudges that help patients gradually reshape their health-related behaviours, such as drinking more water~\cite{Ahire2024}, being more physically active~\cite{Zhang2020, Singh2023}, eating healthily~\cite{Zhang2020, Singh2023} and medical adherence~\cite{Aggarwal2023}.

\begin{table*}[h]
\caption{Summary of Medical Device Regulation (MDR) software classes, including descriptions, example software, required compliance measures (NA = not applicable, TD = Technical Documentation, QMS = Quality Management System), and review processes. CV = Clinical Validation}
\label{MDRClass}
\begin{tabular}{lllll}
\textbf{MDR} & \textbf{Description} & \textbf{Software Example} & \textbf{Compliance} & \textbf{Review} \\ \hline
\textbf{No MDR} & \begin{tabular}[c]{@{}l@{}}Device without an intended\\ medical purpose or clinical use.\end{tabular} & \begin{tabular}[c]{@{}l@{}}General purpose chatbots.\end{tabular} & NA & NA \\ \hline
\textbf{Class I} & \begin{tabular}[c]{@{}l@{}}All software intended for a \\ medical purpose not covered \\ by higher classes.\end{tabular} & \begin{tabular}[c]{@{}l@{}}Fertility status calculator \\ based on validated algorithms.\end{tabular} & TD & Self-certification \\ \hline
\textbf{Class IIa} & \begin{tabular}[c]{@{}l@{}}Software informing clinical-\\ decision making or monitoring \\ physiological processes.\end{tabular} & \begin{tabular}[c]{@{}l@{}}Listing recommended cancer \\ treatment based on patients \\ medical history.\end{tabular} & \begin{tabular}[c]{@{}l@{}}TD \\ QMS \\ CV \end{tabular} & \begin{tabular}[c]{@{}l@{}}Notified Body \\ Limited review \\ of documentation \end{tabular} \\ \hline
\textbf{Class IIb} & \begin{tabular}[c]{@{}l@{}}Software informing decisions\\ which may course serious \\ deterioration of a persons health\\ or monitoring vital physiological\\ parameters.\end{tabular} & \begin{tabular}[c]{@{}l@{}}A mobile app analysing user’s \\ heartbeat, detect abnormalities, \\ and contacting clinicians.\end{tabular} & \begin{tabular}[c]{@{}l@{}}TD \\ QMS \\ CV \end{tabular} & \begin{tabular}[c]{@{}l@{}}Notified Body \\ Expanded review \\ incl. audits\end{tabular} \\ \hline
\textbf{Class III} & \begin{tabular}[c]{@{}l@{}}Software informing decisions \\ which may course irreversible \\ harm or death. Or software \\ providing automatic diagnostic \\ or treatment without clinical \\ oversight.\end{tabular} & \begin{tabular}[c]{@{}l@{}}Image analysis for making \\ treatment decisions in \\ patients with acute stroke.\end{tabular} & \begin{tabular}[c]{@{}l@{}}TD \\ QMS \\ CV \end{tabular} & \begin{tabular}[c]{@{}l@{}}Notified Body \\ Comprehensive review \\ incl. full audits \end{tabular}
\end{tabular}
\end{table*}

\subsection{European regulations}
Designing a chatbot for the European market requires compliance with relevant regulatory frameworks, most notably: Medical Device Regulation (MDR, Regulation (EU) 2017/745~\cite{MDR}), General Data Protection Regulation (GDPR, Regulation (EU) 2016/679~\cite{GDPR}), and the AI Act (AIA, Regulation (EU) 2024/1689~\cite{AIAct}). Together, these govern patient safety, lawful data processing, and the development and deployment of trustworthy AI systems.

\subsubsection{Medical Device Regulation}
MDR requires devices with a medical purpose to demonstrate safety and clinical effectiveness~\cite{MDR_Art5}. Whether a product qualifies for MDR depends on its intended use, as declared by the manufacturer and assessed by regulators~\cite{MDR_Art2}. Thus, seemingly identical products may have different regulatory fates: for example, wooden popsicle sticks are food accessories, whereas identical sticks intended as a tongue depressor during mouth examinations qualify as medical devices. The determining factor is whether the product is intended for a medical purpose, affecting the examination, diagnosis, prevention, monitoring, treatment, or alleviation of an individual's health~\cite{MDR_Art2}. 

MDR distinguishes between Software as a Medical Device (SaMD) and Software in a Medical Device (SiMD)~\cite{MDR_Rec19}. For SiMD, the entire hardware system is regulated as a medical device (e.g., a pacemaker). By contrast, SaMD refers to stand-alone software that qualifies independently; i.e., the host platform (e.g., a smartphone) is not itself a medical device. The remainder of this paper, "software" or equivalent, concerns SaMD. 

Medical device software is classified into four risk-based classes \textemdash~Class I, IIa, IIb, and III \textemdash~with regulatory requirements increasing alongside risk~\cite{MDR_Art51}. Class I devices may rely on literature-based evidence and can be self-certified, provided manufacturers maintain comprehensive technical documentation (TD)~\cite{MDR_Art52}. From Class IIa onwards, manufacturers must also provide clinical evidence and implement a quality management system (QMS), all subject to review by a notified body (NB)~\cite{MDR_Art52}. NB involvement intensifies with risk, ranging from documentation review (IIa) to full audits and continuous oversight (III)~\cite{MDR_AnnIX}. See Table~\ref{MDRClass} for an overview.

MDR defines 22 classification rules, most concerning physical devices~\cite{MDR_AnnVIII}. For software, Rule 11 is typically decisive, although additional rules may apply depending on the software’s intended purpose and functionality. Rule 11 states: Software intended to provide information for diagnostic or therapeutic decision-making is classified as Class IIa. Classification increases when incorrect decisions or software errors can cause serious deterioration in health or require surgical intervention (Class IIb), or result in death or irreversible deterioration (Class III). Software intended to monitor physiological processes is generally Class IIa, but becomes Class IIb when errors in monitoring vital parameters could pose immediate danger. If none of these conditions applies, the software is Class I.

\subsubsection{General Data Protection Regulation}\label{GDPR}
GDPR governs how personal data \textemdash any information that directly or indirectly identifies an individual~\cite{GDPR_Art4} \textemdash is collected, processed, and stored. Obligations scale with data sensitivity. \textit{Anonymous data} (which often needs to be handled as merely highly pseudonymised under the GDPR regime~\cite{GDPR_Art4}) may suffice with basic transparency about who, why, how, and what data is processed. \textit{Personal data} requires a lawful basis for processing~\cite{GDPR_Art6}. \textit{Special category} data~\cite{GDPR_Art9}, such as health data, triggers stricter safeguards and documentation. The entity determining the purpose and means of processing is the \textit{data controller}~\cite{GDPR_Art4} and bears legal responsibility~\cite{GDPR_Art24}. For health chatbots intended for commercial availability, obtaining explicit consent from the individuals whose data is processed is the most appropriate and often the only reliable legal basis~\cite{GDPR_Art9}. 

\subsubsection{The AI Act}
The AIA regulates artificial intelligence through a four-level risk framework~\cite{AIA_Art26}. 1)~Minimal risk AI (e.g. spam filters) face no new obligations (though other laws may apply). 2)~Limited-risk AI (e.g. most chatbots) must meet transparency requirements, including informing users they are interacting with AI~\cite{AIA_Art50}. 3)~High-risk AI (e.g. screening resumes) must implement risk management, bias mitigation, technical documentation, logging, accuracy and cybersecurity safeguards, and human oversight~\cite{AIA_Art26} \textemdash~all subject to review by an NB~\cite{AIA_Art43}. AIs classified as a medical device Class IIa or higher under MDR are automatically deemed high-risk~\cite{AIA_Art50}; in such cases, MDR and AI Act assessments are combined~\cite{MDCG2025, Vardas2025}. 4)~Prohibited AI practices (e.g. manipulative systems) are banned on the EU market~\cite{AIA_Art5}.

\section{Method}
We employed a qualitative, design-oriented analytical approach to examine how conversational design decisions in health chatbots shape regulatory qualification and classification under European frameworks. Using Design Space Analysis, operationalised through the QOC (Questions, Options, Criteria) framework~\cite{MacLean1991}, we conceptualised chatbot behaviour as a conversational design space comprising recurring dimensions (e.g., actionability, proactivity). In QOC, \textit{questions} capture key decisions designers must consider, \textit{options} represent alternative solutions to the questions, and relevant \textit{criteria} evaluate their implications \textemdash~in this study, criteria consisted of regulatory concepts and assessment guidelines. Because each dimension can be realised differently, design choices may lead to different regulatory interpretations.

\subsection{Analytical scope}
The analysis focuses on three conversational functions: 1)~health Q\&A, 2)~recovery monitoring, and 3)~motivating medical adherence.
These functions were selected because they both reflect key post-discharge support needs of stroke survivors (grounding the analysis) and represent informational, monitoring, and persuasive chatbot capabilities that introduce distinct regulatory considerations.

MDR served as the primary analytical anchor, as it governs medical device qualification and classification. GDPR and the AIA additional requirements were included where relevant. To maintain jurisdictional coherence non-EU frameworks were excluded.

\subsection{Analytical procedure}
We applied the QOC framework in four structured stages.

\textbf{Stage 1: Formulation of design questions.} 
Two human–computer interaction (HCI) researchers identified recurring design questions that structure conversational behaviour in health chatbots and capture fundamental design decisions (e.g., intended clinical context). Drawing on prior experience developing a health chatbot for post-discharge stroke care in the Czech Republic, they brainstormed design questions designers must consider when developing healthcare chatbots. Afterwards, they clustered them into recurring themes and refined them into higher-level questions related to health Q\&A, monitoring, and motivational support.

\textbf{Stage 2: Exploration of options.}
For each design question, the HCI researchers identified alternative design options representing different ways of realising the corresponding decision (e.g., conversational roles or levels of actionability). Derived from prior design iterations and development discussions, the final options were selected to capture key variations in conversational behaviour.

\textbf{Stage 3: Apply regulatory criteria.} 
Each option was assessed against MDR, GDPR, and AIA criteria, focusing on its impact on 1)~intended purpose, 2)~clinical authority, and 3)~data type and sensitivity, to determine how design decisions affect regulatory qualification and classification. The HCI researchers conducted the initial assessment to maintain a design perspective.

\textbf{Stage 4: Iterative expert validation.} Four legal scholars independently reviewed the analysis and provided written feedback on potential misinterpretations, ambiguities, and overlooked provisions. Disagreements between the HCI researchers’ proposed regulatory interpretations and the legal experts’ assessments were discussed, where the legal implications of design decisions were clarified. Disagreement among the legal experts themselves was rare and primarily concerned wording rather than overall assessments. This process was repeated three times until a consensus was reached. In the final round, a stroke clinician reviewed the resulting design implications for clinical plausibility. 

The ten design questions and analytical findings are presented in the following sections: Section~\ref{HealthAQ}: \textit{Health Q\&A}, Section~\ref{PROMs}:\textit{ Monitoring recovery}, and Section~\ref{Nudges}: \textit{Motivational Medical Adherence}. To avoid repetition, cross-cutting design questions are addressed only in the section where they have the greatest regulatory relevance.

\subsection{Methodological limitations}
The analytical focus is on design-level implications of regulation rather than broader economic or institutional effects. We therefore do not examine structural barriers, such as administrative burden or cost, that may disproportionately affect small companies, academic groups, or hospital-led innovations~\cite{VanLaere2021}. 

While QOC provides a structured, semi-formal notation for representing design rationale~\cite{MacLean1991}, it does not (necessarily) aim to define a complete set of questions or options. Identifying all possible design questions and corresponding options is inherently impractical \textemdash~if not impossible. The framework should therefore be understood as a structured but non-exhaustive representation of conversational design possibilities and their regulatory implications.

The analysis is situated in the context of post-stroke recovery. While most design implications likely generalise to other clinical contexts, regulatory interpretation is inherently sensitive to the risk profile of intended end-users. Different medical contexts may therefore yield different regulatory assessments.

This analysis is grounded in a European regulatory context, and its findings may not transfer directly to other jurisdictions. Its regulatory interpretation entails inherent subjectivity and uncertainty, as it may vary across regulatory bodies and assessors. Therefore, this paper should not be interpreted as a compliance blueprint. Rather, it illustrates possible interpretive pathways for conversational design. Any specific chatbot implementation should undergo an independent legal review aligned with its intended use and market positioning. 

\section{Analytical findings \textemdash Health Q\&A}\label{HealthAQ} 
A health Q\&A function is often perceived as the most well-suited feature for a health-focused chatbot, as conversational interfaces invite question-response behaviour. However, from a regulatory perspective, even small systematic conversational design decisions may shift a system from delivering generic well-being information to providing patient-specific medical advice. We identified seven design questions that influence the likelihood of a health Q\&A chatbot qualifying as a medical device and subsequent risk classification:

\begin{enumerate}
    \item \textbf{Clinical context:} Is the chatbot targeting a specific patient group or the general public?
    \item \textbf{Scope of inquiry:} What kinds of questions is the chatbot designed to handle?
    \item \textbf{Actionability of responses:} How directly can replies guide user behaviour or treatment decisions?
    \item \textbf{Interaction model:} Does the chatbot only respond to questions, or also proactively provide information?
    \item \textbf{Data sources:} What data are required to generate responses?
    \item \textbf{Answer generation logic:} How are response validity, traceability, and reliability ensured?
    \item \textbf{User self-disclosure:} How does the system handle voluntary disclosure of sensitive health information?
\end{enumerate}

Each question can be realised in multiple ways, and the following subsections show how alternative design options affect regulatory qualification and compliance.

\subsection{Clinical context}\label{Context}
MDR qualification depends not only on functionality, but on the clinical context in which the chatbot operates. Generally, systems targeting specific patient groups or integrated into clinical pathways are more likely to be interpreted as having a medical purpose. For example, a chatbot stating “\textit{Taking a daily walk can improve overall well-being}” would typically fall outside MDR when offered to the general public as wellness advice. However, when intended for stroke survivors and framed as supporting motor recovery, the same recommendation may be considered therapeutic. The behavioural suggestion remains identical, but the disease-specific framing changes its regulatory interpretation. 

For chatbots integrated into clinical routine care or pathways, the prudent assumption is that it qualifies for MDR. For example, a chatbot introduced at hospital discharge and positioned as part of follow-up care will likely be considered as contributing to treatment \textemdash~even if it only provides general information. By contrast, a functionally identical chatbot distributed through an app store without disease-specific claims may fall outside MDR.  

In subsequent (sub)sections, we assume the chatbot is neither disease-specific nor integrated into a clinical pathway unless stated otherwise, as either conditions on their own would typically indicate MDR qualification.

\begin{takeawaybox}
\textbf{Takeaway:} Targeting a specific patient group or clinical pathway increases the likelihood that MDR will apply.
\end{takeawaybox}

\subsection{Scope and clinical role of inquiry}\label{Role}
Although Health Q\&A chatbots appear to perform a single function \textemdash~answering questions \textemdash~their regulatory assessment is more complex. Depending on user queries, the chatbot may provide neutral administrative information in one interaction and influence treatment-related decisions in another. In effect, the system can assume different conversational roles analogous to those of professionals who would traditionally answer such questions. Each role conveys a different level of medical authority and potential impact on health-related decision-making. Because MDR qualification depends on the intended medical purpose~\cite{MDR_AnnVIII}, the level of authority implied by these roles becomes legally significant. For example, answering “\textit{What are common symptoms of a stroke?}” positions the chatbot as a general health educator. By contrast, responding to “\textit{Are my current symptoms a sign of a stroke?}” would require interpreting symptoms and potentially providing medical judgment, posturing the role of a clinician. To illustrate how shifting roles shape regulatory assessment, we analyse four archetypes: \textit{assistant}, \textit{educator}, \textit{coach}, and \textit{clinical expert}.

In this context, 'role' refers to the system’s functional behaviour and outputs rather than its branding and conversational tone. A chatbot marketed as an “AI doctor” may still function as an educator if it provides general information, whereas modestly branded systems may assume a clinical role if they support medical decision-making. Similarly, conversational tone, personality, or anthropomorphic elements (e.g., empathy or friendliness) may influence user trust and engagement but generally do not determine MDR qualification. Regulators primarily assess \textit{what} the chatbot is intended to do, rather than \textit{how} it communicates. However, if conversational tone signals clinical authority, then it may influence how regulators interpret its purpose. 

MDR qualification is assessed based on the chatbot’s intended role rather than isolated responses. Occasional anomalous replies reflecting a higher-risk role do not, in themselves, alter qualification or classification. However, repeated inconsistencies with the stated purpose may indicate mischaracterisation and lead to regulatory reassessment or non-compliance findings. When a chatbot assumes multiple roles, MDR is typically assessed based on the highest-risk intended use. For analytical clarity, each role below is analysed as if it represented the chatbot’s primary intended function.

\begin{takeawaybox}
\textbf{Takeaway:} MDR qualification depends on the chatbot’s overall intended function, assessed by its most medically consequential role, while conversational tone only matters if it signals medical authority.
\end{takeawaybox}

\textbf{Health assistant.} A health assistant chatbot that merely displays or transmits existing factual, logistical, or administrative information, without interpreting it or influencing clinical decisions, is unlikely to be considered a medical device. Even information personalised to the user’s identity or medical record does not alter this assessment. Regardless of MDR qualification, any processing of health data remains subject to GDPR.

\textbf{Health educator.} Transiting from health assistant to educator occurs when the system moves from merely displaying information to explaining it. A chatbot limited to providing general health explanations \textemdash~functioning similarly to a digital FAQ or health library \textemdash~is typically outside MDR since it is not meant to influence treatment. Hence, knowledge-based questions, such as “\textit{What is aphasia?}”, usually prompt descriptive explanations rather than medical guidance. However, if actionable guidance is added, even general information may be considered to have a medical purpose. 

\textbf{Health coach.} A health coach chatbot differs from a health educator by shifting from explaining medical information to encouraging behaviour. In this context, coaching is limited to general health advice and must not aim to influence diagnosis, treatment, or clinical decision-making. Defining what constitutes “general” advice, however, is not always straightforward. Broad public health messages, such as “\textit{Avoid smoking}”, may benefit most people, whereas others, such as “\textit{Engage in physical activity and eat healthy}”, can be inappropriate for certain populations (e.g., individuals with anorexia nervosa). Whether such edge cases affect regulatory qualification depends on the chatbot’s intended purpose. If positioned as providing general wellbeing guidance, risks may be addressed through general disclaimers without requiring individualised assessment. In contrast, if the chatbot is intended to support disease management \textemdash for example, assisting stroke survivors (where anorexia nervosa elevates stroke risk~\cite{Marcolini2024}) \textemdash then contraindications and comorbidities become foreseeable risks that likely trigger MDR qualification.

It may seem that the key distinction lies between general advice (“\textit{30 minutes of exercise a day is good for your well-being}”) and personalised advice (“\textit{To minimise your risk of a second stroke, you should exercise 30 minutes a day}”). However, personalisation alone does not trigger MDR qualification, and even general advice may influence treatment. The decisive factor remains the intended medical purpose of the advice.

\textbf{Clinical expert.} A chatbot posturing as a clinical expert indicates a clearer medical function under MDR~\cite{MDR_Art2}. Unlike a health coach providing general wellness encouragement, experts engage with disease-specific, treatment-oriented, and potentially safety-critical inquiries. For example, answering “\textit{Can I drink wine while on my stroke medication?}” requires interpreting drug interactions, treatment regimens, and possible contraindications. Such responses go beyond education and directly influence treatment adherence and patient decision-making. Even when based on established guidelines or drug databases, applying this knowledge to a user’s therapeutic context involves clinical interpretation rather than simple information retrieval. Hence, clinical expert chatbots are generally classified as Class IIa or higher.

\textbf{Summary.} These four roles illustrate a gradient of conversational authority with increasing influence on health-related decisions. This shift is not merely semantic but affects the likelihood that a chatbot qualifies as a medical device. Designers must therefore explicitly define the maximum clinical authority their chatbot is permitted to exercise, as this upper boundary will ultimately contribute to its legal classification and compliance obligations.

\begin{takeawaybox}
\textbf{Takeaway:} Chatbots with greater clinical authority are more likely to qualify as medical devices.
\end{takeawaybox}

\subsection{Degree of actionability in chatbot replies}~\label{actionability}
Beyond the chatbot’s role and the type of inquiry it enables, MDR qualification is strongly influenced by the actionability of its responses. The more directly a chatbot’s output guides behaviour related to diagnosis, monitoring, or treatment, the more likely it is to indicate a medical intended purpose. In this sense, seemingly informational features can \textit{qualify} within MDR scope, with \textit{classification} depending on the level of actionability and the potential severity of harm.

\subsubsection{Actionability influence on qualification.} 
Chatbot roles that could avoid MDR can trigger qualification by increasing actionability. Consider the difference between an educator stating, “\textit{Facial drooping and arm weakness may indicate a stroke}” versus, “\textit{If you are experiencing facial drooping and arm weakness, you may be having a stroke and should seek emergency care.}” Both statements rely on the same general medical knowledge, but the latter frames the information as directly applicable to the user and recommends a course of action. If the chatbot is designed to assess reported symptoms and guide treatment-seeking behaviour, it may effectively perform a triage function, thereby indicating a medical purpose. Importantly, encouraging users to consult a clinician alone does not constitute a medical purpose, and generic signposting is often encouraged to ensure patient safety. What moves a chatbot toward MDR qualification is the systematic evaluation of user-specific information and the generation of case-specific recommendations.

Similarly, an assistive chatbot replying to the following inquiry: “\textit{I’m supposed to take anticoagulants in the morning, right?}” could either 1)~retrieve stored information ("\textit{According to your medical record you are supposed to take anticoagulants at noon}"), or 2)~interpret and confirm treatment correctness ("\textit{No, you are supposed to take that at noon}"). While both rely on the same data, confirming or correcting treatment behaviour influences therapy management and may therefore support medical device qualification.

\begin{takeawaybox}
\textbf{Takeaway:} Increasing the actionability of chatbot replies (e.g. going from informing and explaining to recommending or verifying), roles otherwise outside MDR scope might qualify.   
\end{takeawaybox}

\subsubsection{Actionability influence on classification.} 
Once a chatbot qualifies as a medical device, the actionability of its responses influences its MDR risk classification. If qualification arises from contextual factors elsewhere in the system (e.g., integration into a clinical pathway), low-actionability functions, such as administrative support or general education, may be defensible as Class I, provided they do not influence diagnosis, monitoring, or treatment.

When qualification stems from the actionability of the chatbot’s replies, the system typically supports clinical decision-making and is therefore classified as at least Class IIa. Final classification depends on the potential consequences of erroneous outputs. Patient-facing systems that provide automated diagnostic or therapeutic guidance, or predictive health modelling, may be considered higher risk, as users are not expected to have the medical expertise to critically evaluate the advice. The inclusion of human oversight may mitigate risk, but does not automatically reduce classification \textemdash~Class IIa remains the lowest class for a system supporting diagnostic or therapeutic decisions~\cite{MDR_AnnVIII}. However, meaningful human reviews may shift the system from driving decisions to merely supporting them, thereby influencing whether classification remains at Class IIa or escalates to higher classes.

\begin{takeawaybox}
\textbf{Takeaway:} Chatbots qualifying for their actionability typically support diagnostic or therapeutic decisions, making them at least Class IIa, even when including human oversight.   
\end{takeawaybox}

\begin{table*}[h]
\centering
\caption{Different proactive approaches affect the perceived likelihood that the chatbot performs a medical function. This assumes that the general health-related advice coach is not part of a clinical pathway as that might trigger MDR regardless of proactivity.}\label{Proactivity}
\begin{tabular}{p{0.13\linewidth} p{0.25\linewidth} p{0.25\linewidth} p{0.3\linewidth}}
\textbf{Advice trigger} & \textbf{Description} & \textbf{Example} & \textbf{Likelihood of MDR} \\
\hline
No trigger & 
The same generic advice is shown to all users without personalisation or a trigger. & 
“\textit{Tip of the day: Staying physically active supports general wellbeing.}” & 
Low likelihood. \\
\hline
Event-based & 
Advice triggered by time, seasons, or generic milestones. & 
“\textit{Winter is approaching; respiratory infections increase during this season.}” &
Moderate likelihood.\newline  
May indicate a medical purpose if events correspond to recognised treatment milestones. \\
\hline
Topic-based trigger & 
Advice triggered by topics mentioned in conversation but not based on personal health data. & 
User mentions travelling → “\textit{Pack medication in separate bags when travelling.}” &
Low to moderate likelihood.\newline 
Topic dependent; health topics may increase perceived clinical relevance. \\
\hline
Interaction-based tailoring & 
The chatbot asks follow-up questions to gather context and adapt its advice. & 
“\textit{Are you planning to travel abroad?}” → yes → “\textit{If travelling with medication, bring a copy of your prescription.}” &
Low to moderate likelihood. \newline 
Regulatory relevance depends on whether collected context is used to guide diagnosis, monitoring, or treatment.\\
\hline
Condition-based tailoring  & 
Advice personalised using known health conditions or medical history. & 
“\textit{Since you have high blood pressure, remember to take anticoagulants.}” &
High likelihood. \newline
Applying medical knowledge to a user’s health condition typically indicates a medical purpose. \\
\hline
Predictive tailoring & 
The system anticipates needs using behavioural patterns or health data. & 
“\textit{You may be running low on your prescription—consider requesting a refill}.” &
High likelihood. \newline
Predicting users health needs without human oversight present a higher risk.  \\
\end{tabular}
\end{table*}

\subsection{Reactive versus proactive interaction}
While the actionability of a chatbot’s responses determines how strongly its outputs guide user behaviour, another design decision concerns \textit{when} such guidance is provided. Health Q\&A systems can be purely reactive, responding only to user queries, or incorporate proactive features that surface information that users might not otherwise request. For example, advice such as “\textit{When travelling, pack your medicine in multiple bags, including carry-ons, that way you don’t lose all your medication if you should lose a bag}” may be useful yet unlikely to be prompted by a user question. Providing such unsolicited general well-being advice does not, by itself, trigger MDR qualification; many wellness chatbots offer similar reminders without being considered medical devices~\cite{MDR_Rec19}. However, proactive features may matter in borderline cases because they can strengthen the perception that the system is intended to guide health behaviour. This is particularly relevant when unsolicited advice is targeted at a specific patient group or embedded within a clinical pathway. In such contexts, proactive guidance may contribute to the impression that the chatbot has a medical intent. The regulatory implications, therefore, depend on how and why unsolicited advice is triggered (see Table~\ref{Proactivity}). 

\begin{takeawaybox}
\textbf{Takeaway:} Proactive advice alone does not trigger MDR, but it can reinforce the perception that the chatbot is guiding health decisions.
\end{takeawaybox}

\subsection{Answer generation logic}
When designing a health Q\&A chatbot, developers must choose how user questions are interpreted and answers generated. Three broad approaches are common: rule-based scripting, generative large language models (LLMs), and retrieval-augmented generation (RAG). While appearing similar to users, these approaches differ substantially in controllability, explainability, and validation. MDR qualification depends on whether the chatbot has a medical purpose, not on its technical architecture. However, once the system qualifies as a medical device, the answer-generation logic strongly affects the feasibility and cost of compliance, as manufacturers must demonstrate safety, performance, and robust risk management.

\textbf{Rule-based chatbots} rely on predefined decision trees that map recognised inputs to validated outputs. This offers high controllability and traceability: all responses can be reviewed, documented, and clinically validated in advance. While compliance obligations remain, rule-based systems simplify technical documentation compared to more autonomous designs. The trade-off is limited conversational flexibility and extensive scripting to handle linguistic variation. Crucially, purely rule-based chatbots \textemdash~using only if-then logic, decision trees, with no learning or adaptive inference \textemdash~are not considered AI under the AIA and are exempt from its requirements even if the chatbot is a medical device. Any addition of NLP, machine learning, or adaptive capabilities would trigger AIA obligations.  

\textbf{Generative LLM-based chatbots} produce responses dynamically using probabilistic language models, enabling flexible interaction and broad query coverage. However, outputs are not predetermined, making full validation complex. Variability and the risk of inaccurate or fabricated information increase the burden on manufacturers to demonstrate consistent safety and performance within the intended clinical context. While LLM-based chatbots are not prohibited under MDR, their autonomous nature requires extensive validation. To our knowledge, no purely generative model has obtained MDR classification. 

\textbf{Retrieval-augmented generation} combines flexibility and control by retrieving the most relevant answer from a curated, pre-approved repository of validated documents. An LLM optionally assists with semantic matching, summarisation, or paraphrasing. This simplifies technical documentation as each response can be traced back to source documents, but the retrieval and matching mechanism must be validated to ensure accuracy. Using RAG-based knowledge retrieval from a controlled database, "Prof. Valmed" became the first Class IIb-certified chatbot for diagnostic and therapeutic decision support~\cite{ProfValmed2025}. Demonstrating RAG can support higher-risk certification, although regulatory approval depends on the specific implementation and evidence provided. 

\begin{takeawaybox}
\textbf{Takeaway:} Once a chatbot qualifies as a medical device, its answer generation logic determines traceability, validation effort, and compliance burden.
\end{takeawaybox}

\subsection{Data sources}
When answering questions, the chatbot requires access to relevant information, which varies with the inquiry. For example, answering "\textit{What is a stroke}?" only requires general information, whereas "\textit{Which stroke did \textbf{I} have}?" needs the user's medical history. MDR qualification depends on how the data is used, not its source. Hence, a health assistant may display medical record information without being a medical device, whereas a coach that relies solely on general knowledge may qualify if its responses are intended to support medical decision-making.

The type of data processed is critical under GDPR. Truly anony\-mised data fall outside its scope, but this is difficult for most chatbots, which typically handle metadata \textemdash~IP addresses, device IDs, timestamps, or account details \textemdash~which are considered personal data. Therefore, it is prudent to assume the chatbot processes personal data. Extra compliance requirements apply only to special-category data, such as health information~\cite{GDPR_Art9}. Importantly, health information is only considered a special category when it relates to an identifiable individual~\cite{GDPR_Art4}. Hence, an educator chatbot that provides encyclopedia-style health knowledge does not necessarily handle special-category data. 

One type of information that avoids both MDR and extra GDPR obligations, but still requires careful consideration, is legal information. Some health-related questions intersect with local laws. For example, “\textit{When can I drive again after my stroke?}” depends on both the individual’s medical history and local driving regulations. Under the AIA, a chatbot providing legal advice that is tailored to individual circumstances or used to inform legally significant decisions would likely be classified as high-risk AI~\cite{AIA_Art6}. High-risk AI systems must meet stringent requirements (similar to those of Class IIa medical devices), including risk management, transparency, accuracy validation, human oversight, and, in some cases, review by a notified body. General legal information that does not apply to specific users or influence decisions is lower risk but should still be clearly framed as informational. 

\begin{takeawaybox}
\textbf{Takeaway:} MDR qualification does not depend on the data source. GDPR likely applies with extra obligations for special category data. Legal data requires careful handling, as tailored advice can trigger high-risk classification under AIA. 
\end{takeawaybox}

\subsection{User disclosing sensitive information}\label{Self-Disclose}
Chatbots typically accept free-text input, giving manufacturers minimal control over what users disclose~\cite{Hasal2021}. As a result, users may voluntarily reveal sensitive information such as health status, medications, or personal details~\cite{Cox2025}. For example, a user might prompt "\textit{I'm 35 years old and had an ischemic stroke, is that normal?}". Regardless of the developer’s intentions, receiving, storing (e.g., in logs), or automatically detecting such self-disclosed content constitutes processing health data. Under GDPR, this means the developer becomes a data controller with full legal responsibilities. Because such self-disclosure is foreseeable in open-ended chat interfaces, users should be informed of the risks and the legal basis for processing \textemdash~ideally, explicit consent \textemdash~should be obtained before using the chatbot. 

Unprompted self-disclosure places chatbots in an MDR grey zone. While general-purpose systems (e.g. ChatGPT) also risk user self-disclosure, they typically fall outside MDR because they are not intended for medical purposes and include disclaimers warning against relying on them for health advice. In contrast, health-focused Q\&A chatbots invite health-related interaction, making self-disclosure foreseeable. If developers can guarantee that disclosed information is ignored and not used to shape responses, MDR qualification may remain unchanged. However, if the chatbot uses the information to personalise replies or guide decisions, this can indicate a medical purpose and trigger MDR compliance. Consequently, even chatbots intended to provide general information may require MDR oversight if they cannot reliably prevent such data from influencing their replies.

\begin{takeawaybox}
\textbf{Takeaway:} Users may disclose sensitive health information in open-text chatbots. Any processing of health data triggers GDPR, and using it to personalise advice can trigger MDR. 
\end{takeawaybox}


\section{Analytical findings \textemdash Monitoring recovery}\label{PROMs}
Monitoring patient recovery helps clinicians adapt care and supports institutional quality improvement. Although many types of health data can be collected (e.g., physiological measures, symptom diaries, PROMs), stand-alone chatbots typically rely on manual user input. Their regulatory status, therefore, depends mainly on how the chatbot uses these data, with an important exception when measures contain high-risk indicators (e.g., suicidal ideation) that should trigger safety mechanisms and shape regulatory assessment. This leads to two additional design questions:

\begin{itemize}
    \item[(8)] \textbf{Data purpose:} What role does the collected data play in informing clinical decision-making?
    \item[(9)] \textbf{Safety handling:} How should the system detect and respond to high-risk disclosures?
\end{itemize}

\subsection{Role of data in clinical decision}
Traditionally, paper-based PROMs were treated as survey instruments rather than medical devices because clinicians performed all scoring and interpretation. Chatbots can automate these tasks and may therefore qualify as medical devices under MDR, but monitoring recovery does not automatically trigger MDR. Instead, qualification depends on how PROM data are used and on the chatbot’s analytical capabilities. To illustrate this, we consider three scenarios: 1)~the chatbot only collects PROM data for a health registry, 2)~PROM data are forwarded to clinicians to inform treatment decisions, and 3)~the chatbot presents a summary of PROM results directly to the patient.

\textbf{Scenario 1: PROMs for a health registry.}  
A key factor of MDR is that it applies only to software intended to influence an individual patient’s care. Chatbots collecting PROMs solely for registry research, public health, or quality improvement remain outside MDR, even if the data are identifiable, as long as the analysis is aggregate and does not affect individual treatment. Likewise, non-re-identifiable PROM collection always remains outside MDR, as it can only generate population-level insights.

\textbf{Scenario 2: Informing clinicians.}
When PROMs are used within an individual patient’s clinical pathway, MDR applies, and classification depends on the level of automatic analysis. A chatbot that only collects responses \textemdash without interpretation \textemdash serves as a data-entry tool and is typically Class I. Convenience features such as summarising scores via standard algorithms likely remain Class I as long as clinicians can verify using the raw data, but unverifiable scores likely raise risk to Class IIa. Automatic score interpretation constitutes a medical function, with classification varying between Class I and  IIa depending on interpretation complexity and data transparency. If the chatbot suggests a diagnosis or treatment based on its interpretation, it directly supports clinical decisions and is generally Class IIa, escalating to Class IIb for high-risk areas such as suicide or life-threatening conditions.

\textbf{Scenario 3: Patient-facing PROMs.}
Clinicians can interpret PROM results within a broader clinical context, whereas patients may misinterpret or overreact, increasing potential harm. Consequently, patient-facing PROM chatbots are generally higher risk than clinician-facing ones. Collecting and displaying responses back to the patient without analysis typically remains Class I. However, providing scores or interpretations typically elevates the device to Class IIa or higher, while automatically suggesting diagnoses or treatments without clinical oversight may warrant Class III for suicide or other life-threatening conditions.

\begin{takeawaybox}
\textbf{Takeaway:} Collecting PROMs alone mostly carries low risk, but directing them to patients or adding scoring, interpretation, or treatment guidance increases MDR classification.
\end{takeawaybox}

\subsection{Handling high-risk indicators}~\label{SelfHarm}
Multiple PROMs measure the patient's mental health, which may include items on self-harm or suicide. For example, the Patient Health Questionnaire-9 (PHQ-9): "\textit{During the past two weeks, how often have you had thoughts that you would be better off dead or of hurting yourself in some way?}"~\cite{Williams2005}. Chatbots that collect such information may require built-in safeguards to ensure patient safety. 

Historically, at-home PROMs were distributed and returned by post, taking weeks to reach clinicians; therefore, there was no expectation of immediate support. Thus, the absence of safeguards was understood as a limitation of the medium, not a failure of duty. Chatbots shift this expectation. Because they detect and respond instantly, failing to address disclosures of self-harm may be seen as neglect or indifference. Under MDR, manufacturers must identify foreseeable risks and implement mitigations~\cite{MDR_AnnIX}. Since PROMs directly ask about suicidal intent, it becomes a foreseeable harm, and ignoring it could constitute non-compliance. Even chatbots that are normally outside MDR, such as those collecting PROMs for aggregate analysis, may still be liable under national tort laws, as most EU member states recognise a duty to prevent foreseeable harm. To be legally robust, chatbots collecting PROMs with self-harm items should implement safeguards. Some safeguards, however, may bring the software under MDR, even if the PROM collection alone would not. This creates a dilemma: ignore disclosures and risk liability, or add safeguards and be classified as a medical device. The type of safeguard also affects classification. We considered four approaches: 1)~excluding PROMs concerning self-harm, 2)~signposting to suicide prevention resources, 3)~automatically alerting clinicians, and 4)~acting as a crisis support service.

\textbf{Actively excluding PROM items}: Removing suicide-related questions avoids safeguard obligations. However, if the PROM collection is intended for a specific patient group with elevated suicide risk \textemdash~such as stroke survivors~\cite{Eriksson2015, Pompili2012} \textemdash~deliberately ignoring these issues may constitute negligence of foreseeable harm and create a critical clinical blind spot \textemdash~this may not be the case for other patient groups. While this may be acceptable in research or development, it is unlikely to be permissible if the chatbot is recommended, prescribed, or used within a healthcare setting. 

\textbf{Signposting}: The act of directing users to external resources or actions without providing personalised advice or making decisions on their behalf. For example, providing a standard message following a sensitive PROM item (e.g. “\textit{If you are having thoughts of harming yourself, please consider calling...}") preserves patients' autonomy and does not constitute a medical function, so it generally does not trigger MDR. 

\textbf{Automatic clinician alerts}: If a chatbot automatically notifies a healthcare provider, it directly supports clinical decision-making, qualifying as a medical device \textemdash~likely Class IIb for monitoring critical health parameters. Consequently, PROM-collection chatbots intended only for aggregated analysis would suddenly fall under MDR, and previously low-risk chatbots would likely be reclassified. In anonymous PROM collection, automatic alerts are technically impossible as the patient cannot be identified. 

Automatic alerts also raise ethical concerns. While intended to protect patients, they may undermine their autonomy by removing the choice to seek help, conflicting with the principle that competent patients can refuse treatment even at serious risk~\cite{Burr2020, Varkey2020}. One mitigation strategy is to request the patient’s consent before alerting clinicians, thereby reducing practical barriers to accessing care while preserving choice. However, transparency about such alert systems may produce unintended consequences; some users might trigger alerts deliberately to access services faster, while others may withhold disclosures due to stigma, fear of judgment, or concerns about coercive intervention~\cite{Eigenhuis2021, Hoyen2022}.

\textbf{Active counselling or therapy}: Chatbots that autonomously provide crisis counselling or therapeutic intervention in potentially life-threatening situations may fall into the highest-risk category (Class III) if errors could result in death or irreversible harm~\cite{MDR_AnnVIII}. The escalation in classification reflects a functional shift from passive intermediary to stand-alone therapeutic actor. While such systems don’t require generative models, most contemporary chatbots would likely rely on LLMs. It remains uncertain whether current LLM models can meet regulatory demands for low-risk classifications, let alone comply with the strict requirements for high-risk classifications. The safety, ethical, and regulatory challenges of deploying chatbots in crisis contexts continue to be debated~\cite{Maples2024, Sedlakova2023, Meadi2025}.

\begin{takeawaybox}
\textbf{Takeaway:} MDR compliance requires safeguards to foreseeable risks, but the chosen option strongly influences whether the chatbot qualifies as a medical device and its class.
\end{takeawaybox}


\begin{table*}[h]
\centering
\caption{Ethical principles illustrated through nudges versus manipulations (leading to exclusion under the AI Act).}
\label{nudges}
\begin{tabular}{llll}
\textbf{Principle} & \textbf{Description} & \textbf{Nudge} & \textbf{Manipulation} \\ \hline
\textbf{\begin{tabular}[c]{@{}l@{}}Respecting \\ autonomy\end{tabular}} & \begin{tabular}[c]{@{}l@{}}Nudges should remain \\ optional and preserve \\ user choice.\end{tabular} & 
\textit{\begin{tabular}[c]{@{}l@{}}“It’s time for your hand \\ exercises, would you like \\ to start now or later?”\end{tabular}} & 
\textit{\begin{tabular}[c]{@{}l@{}}“Start your hand exercises now \\ or your recovery will fail.”\end{tabular}} \\ \hline
\textbf{Transparency}                                                           & \begin{tabular}[c]{@{}l@{}}Suggestions should be \\ clear and their purpose \\ explained.\end{tabular} & 
\textit{\begin{tabular}[c]{@{}l@{}}“Based on your rehab plan,  \\ it’s time for your daily hand \\ exercises.”\end{tabular}} &
\begin{tabular}[c]{@{}l@{}}\textit{“You should do your exercises} \\ \textit{now.”} (without explanation)\end{tabular} \\ \hline
\textbf{Truthfulness}                                                           & \begin{tabular}[c]{@{}l@{}}Nudges must not use \\ fabricated claims e.g. \\ social proof or urgency.\end{tabular} &                                                                         \textit{\begin{tabular}[c]{@{}l@{}}"Answering today’s PROM \\ will help track your recovery \\ progress."\end{tabular}} & 
\textit{\begin{tabular}[c]{@{}l@{}}“90\% of patients like you \\ never skip their exercise.”\end{tabular}} \\ \hline
\textbf{Proportionality}                                                        & \begin{tabular}[c]{@{}l@{}}Expectations should \\ reflect real benefits \\ and risks.\end{tabular}  & 
\textit{\begin{tabular}[c]{@{}l@{}}“Taking your medication \\ supports your recovery.”\end{tabular}} & 
\textit{\begin{tabular}[c]{@{}l@{}}“Taking your medication \\ ensures you will never have  \\ another stroke.”\end{tabular}} \\ \hline
\textbf{\begin{tabular}[c]{@{}l@{}}Considering \\ vulnerabilities\end{tabular}} & \begin{tabular}[c]{@{}l@{}}Nudges must not exploit \\ emotional states, such as\\ fear, or guilt.\end{tabular}  &   
\textit{\begin{tabular}[c]{@{}l@{}}"If you’re feeling tired, you \\ could try a shorter version \\ of your exercises today."\end{tabular}} & 
\textit{\begin{tabular}[c]{@{}l@{}}“If you skip exercises, \\ you are failing your family.”\end{tabular}} \\ \hline
\textbf{\begin{tabular}[c]{@{}l@{}}Alignment with \\ user goals\end{tabular}}   & \begin{tabular}[c]{@{}l@{}}Nudges should support \\ the user’s stated objectives.\end{tabular} & 
\textit{\begin{tabular}[c]{@{}l@{}}“You set a goal of walking \\ daily, want to go now?”\end{tabular}} &
\textit{\begin{tabular}[c]{@{}l@{}}“You should be walking 10,000 \\ steps a day, get going.”\end{tabular}     }                                          
\end{tabular}
\end{table*}

\section{Analytical findings \textemdash Motivational medical adherence}\label{Nudges}
Supporting patients in adhering to prescribed medication and treatment plans is a persistent clinical challenge, as non-adherence can reduce therapeutic effectiveness, increase complications, and generate avoidable healthcare costs. Chatbots can help by providing reminders, encouragement, and behavioural nudges in an interactive format. A motivational chatbot differs from a health Q\&A coach by proactively influencing what the patient does, rather than merely what the patient knows or receives as recommendations. However, like health advice, behavioural nudges often occupy a regulatory grey zone. While behavioural guidance alone does not qualify a chatbot as a medical device \textemdash the key factor remains the intended medical purpose \textemdash~more targeted nudging that moves from generic prompting to tailored behavioural management increasingly resembles treatment support or therapeutic monitoring, raising the likelihood of MDR qualification. This raises a final design question:

\begin{itemize}
    \item[(10)] \textbf{Nudge threshold:} At what point does motivational support become manipulative influence?
\end{itemize}

\subsection{The fine line between nudging and manipulating}
For the health Q\&A and recovery monitoring, the AIA risk assessment largely mirrored the MDR analysis. Chatbots that qualify as Class IIa or higher medical devices are also considered high-risk AI, whereas general well-being chatbots outside MDR or classified as Class I typically fall into the limited-risk category~\cite{AIA_Art6}. Individual design variations within these functions were unlikely to change the overall AIA risk assessment \textemdash~except when providing legal advice. Motivational features introduce a distinct AIA dimension that may ultimately determine whether a system is permitted.

The AI Act prohibits AI systems that use subliminal techniques, deception, or exploit vulnerabilities in a manner that materially distorts behaviour and is likely to cause significant harm~\cite{AIA_Art5}. A chatbot that meaningfully impairs a patient’s autonomous decision-making \textemdash~especially by exploiting health-related vulnerabilities \textemdash~could therefore fall under these prohibited practices. What counts as harmful manipulation is context-sensitive and depends on the type of influence, the vulnerability of the user group, and the likelihood of harm. For instance, in high-stakes domains such as finance, even minor manipulations may meet this threshold.

Motivational chatbots, designed to improve users’ well-being and with a clear medical purpose, might more plausibly claim that their nudges serve the patient's best interest. However, it is not a free pass, and motivations should respect key ethical principles: user autonomy, transparency, truthfulness, proportionality, sensitivity to vulnerabilities, and alignment with user goals (see Table~\ref{nudges}). When these principles are followed, common interventions such as reminders, positive reinforcement, or personalised suggestions are generally permissible.

Nudges must avoid conflicts of interest, particularly regarding medications or commercial products. Encouraging adherence to a clinician-prescribed treatment is generally compatible with MDR and AIA if it is not manipulative. In contrast, nudging users toward specific brands or commercial products may raise issues under consumer protection law, including the \textit{Directive on medicinal products} (EU 2001/83/EC~\cite{Directive2001}), which restricts direct-to-consumer promotion of prescription drugs, and the \textit{Unfair Commercial Practices Directive} (EU 2005/29/EC~\cite{Directive2005}) which bans hidden or misleading commercial influence.

\begin{takeawaybox}
\textbf{Takeaway:} Behavioural nudges are lawful when transparent and autonomy-respecting, but manipulative or exploitative influence \textemdash~especially targeting health vulnerabilities \textemdash~can render a chatbot prohibited.
\end{takeawaybox}

\section{Discussion}
Our findings show that conversational interaction design in healthcare chatbots is not merely a usability concern, but directly shapes regulatory interpretation. Importantly, this relationship cannot be understood through a purely technology-first approach that assumes identical technologies will always receive the same regulatory assessment~\cite{Haupt2023}. For example, image processing is common in non-medical applications, yet the same capability may qualify as a medical device when used for cancer screening or diagnostic support. Under MDR, qualification depends not on technology alone, but on the sociotechnical context, including how systems are framed, positioned, and intended for clinical use.

This means conversational features such as personalised guidance, recommendations, or safeguards are not inherently medical. Rather, regulatory qualification emerges from their combination with intended purpose, target users, and deployment context. We therefore conceptualise healthcare chatbots as recurring design questions with alternative options, enabling analysis across configurations rather than single implementations. This abstraction improves transferability across technologies and contexts. While grounded in stroke recovery, the identified questions apply broadly across healthcare chatbot design.

\begin{figure*}[h]
  \includegraphics[width=0.8\textwidth]{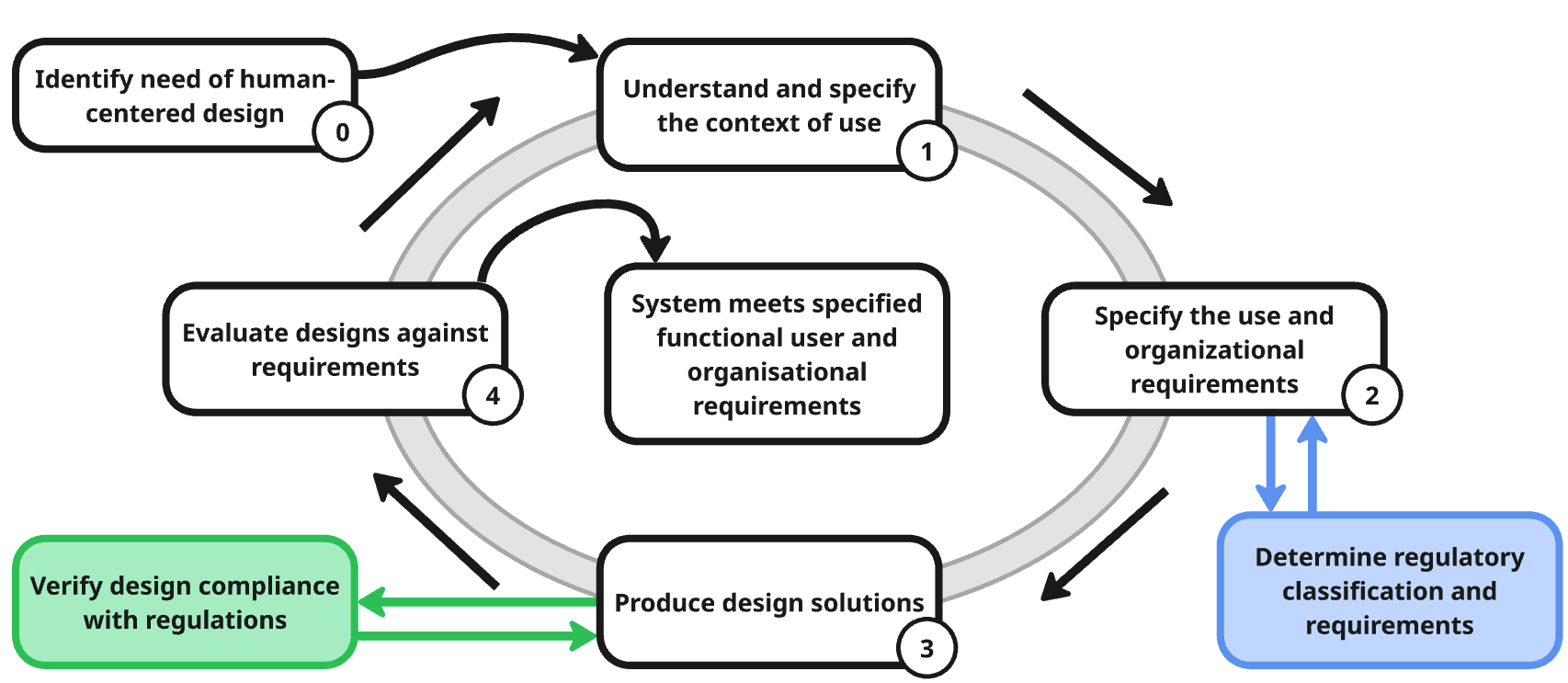}
  \caption{The Human-Centred Design framework enriched with regulatory considerations through legal consultation after: 1)~specifying requirements, to determine appropriate regulatory classification; and 2)~producing design solutions, to verify compliance and ensure that no additional approvals are inadvertently required.}\label{DesignWorkflow}
  \Description{The diagram illustrates the Human-Centred Design (HCD) process as a cycle of five main stages: 0. Identify the need for human-centred design. 1. Understand and specify the context of use. 2. Specify the user and organisational requirements. 3. Produce design solutions 4. Evaluate designs against requirements. These stages are shown as connected boxes with arrows forming an iterative loop. At the centre is a box reading “System meets specified functional user and organisational requirements.” Two additional coloured boxes represent regulatory checkpoints: A blue box connected to stage 2 (“Specify the user and organisational requirements”). It reads “Determine regulatory classification and requirements”, meaning designers should consult legal experts after the requirements are defined to identify the regulatory needs. A green box connected to stage 3 (“Produce design solutions”). It reads “Verify design compliance with regulations”, meaning designers should ensure their design remains compliant and does not trigger a higher classification. Together, these additions show how regulatory steps can be integrated into the HCD workflow for healthcare contexts.
  }
\end{figure*}

\subsection{The tension between HCD and regulation}
Human-centred design (HCD) typically focuses on how features support users rather than on their regulatory implications. For example, a health Q\&A chatbot may integrate generative AI with patient records to provide personalised recommendations. However, increasing personalisation, actionability, and clinical authority raises regulatory obligations. This creates a tension between human-centred and regulatory-centred design. While the former prioritises user experience, frameworks such as MDR emphasise risk mitigation, accountability, and clear clinical boundaries. Designers may therefore face trade-offs between expanding functionality to better support users and limiting capabilities to maintain a lower-risk classification. From an HCD perspective, constraining beneficial functionality may appear counterintuitive, yet it may be necessary for safe and deployable systems.

This tension is particularly evident in research contexts, where prototypes often operate under ethical approval rather than medical regulation. Although this enables exploratory innovation, it can create a disconnect between promising results and real-world deployability. Regulatory considerations should therefore be treated as design constraints throughout development. Otherwise, features introduced during prototyping may later prove incompatible with the intended regulatory pathway, requiring substantial redesign.

\subsection{Integrating legal assessment into design workflows}
Traditional HCD frameworks generally treat “legality” as implicit or external to the design process, leaving designers insufficiently equipped to anticipate how design decisions affect qualification and classification~\cite{dickhaut2022role, Monteiro2026}. To mitigate the risks of ad-hoc regulatory compliance, we advocate that healthcare-focused HCI designers integrate legal reviews into their workflows. Following the human-centred design framework~\cite{ISO_HCD}, we recommend consulting legal experts at least twice during development (see Figure~\ref{DesignWorkflow}).

First, after specifying user and organisational requirements, identify feasible regulatory pathways and involve relevant stakeholders to determine the appropriate MDR classification level (if any), thereby guiding the design process. At this stage, QOC, or similar design-space frameworks, paired with expert validation, provided four methodological advantages. 1)~It transformed regulatory ambiguity into concrete conversational design questions. 2)~Alternative design options enabled designers and stakeholders to reason about legal trade-offs rather than binary compliance outcomes typically associated with assessing a single system. 3)~It enabled interdisciplinary dialogue between HCD and legal experts by grounding legal interpretation in concrete conversational behaviours and design alternatives. 4)~Framing the analysis as recurring design questions supports transferability across healthcare features and domains where similar conversational considerations arise.

Second, after producing a design solution, verify that no new design feature inadvertently alters qualification or classification. Conducting this assessment prior to evaluation can prevent the use of testing systems that may not adhere to regulatory requirements, saving significant time and money. The goal is to cultivate a cooperative design process in which clinicians, designers, and legal experts integrate both human needs and regulatory obligations.

\subsection{Who should aim for the last mile?}
For certain AI systems \textemdash~such as cancer screening and radiology diagnostics \textemdash~clinical deployment and medical device qualification are unavoidable because their intended purpose is clearly medical. Patient-facing chatbots, however, often occupy a regulatory grey zone. Depending on how their intended purpose is defined and operationalised, they may qualify either as medical devices under the MDR or as general well-being applications. In these cases, transitioning to a medical device becomes a deliberate design decision, raising the broader question: should designers always strive for the 'last mile' of clinical care, or could it sometimes be preferable to 'turn left' toward general well-being applications?

Qualifying as a medical device offers important advantages. Beyond enabling treatment-supporting functionality, MDR certification can increase trust among clinicians and healthcare institutions, making them more likely to integrate the system into clinical pathways and thereby provide systematic access to target patients. The trade-off is the associated financial and time costs, especially for higher classifications, which require NB involvement, clinical evaluation, and ongoing conformity procedures. 

Remaining outside MDR, as a well-being application, reduces regulatory burden (saving money and time) and allows greater flexibility in conversational design and AI capabilities \textemdash~although GDPR and AI Act obligations still apply. However, such systems must avoid medical claims and individual clinical decision support, which may limit clinical integration and the associated benefits. Reaching specific patient groups may require careful communication strategies that avoid unintentionally implying a medical purpose. Nevertheless, meaningful health impact does not necessarily require medical device qualification; many tools support health-related behaviours without qualifying as medical devices. For example, a standard smartphone alarm can effectively assist with medication adherence by setting reminders.

Developers must therefore weigh the intended impact, clinical integration goals, and regulatory costs. In some cases, remaining within the well-being domain may be both strategic and ethically appropriate; in others, meaningful clinical influence justifies the additional costs associated with crossing the 'last mile'.

\subsection{Approaches for going to market}
If MDR certification is needed, one might intuitively assume the only option is to design for, and apply directly to, the final classification. While this is true for most physical medical devices\textemdash~a pacemaker cannot be classified below Class III\textemdash~this is not necessarily the case for software, which can evolve significantly after launch. For software, this opens up two mutually inclusive pathways: incremental classification and architectural modularisation. 

\textbf{Incremental Classification}: While medical device qualification depends on the intended medical purpose, its subsequent classification is determined through a more rule-based risk assessment: software that influences diagnostic or therapeutic decisions is typically classified as Class IIa or higher, depending on the severity of potential harm. Designers are therefore flexible in moving between classes by altering functionality to be more or less risky. 

Where feasible, developers may therefore begin with systems that deploy lower-risk functionality before transitioning to Class IIa or higher features. This enables an incremental, regulatory-aware product roadmap in which an initial version is launched early to gather real-world feedback from patients and clinicians, while more advanced decision-support capabilities are introduced later. Where classification as Class I is possible, self-certification supports a more iterative HCD process, as new or improved features generally only require updates to the technical documentation \textemdash~provided the intended purpose remains unchanged, and reclassification is not triggered. By contrast, substantial modifications to Class IIa or above systems typically require renewed NB review and additional clinical evidence, making changes both slower and significantly more expensive.

The rare case, a Class I AI system is deemed high-risk under AIA independent of MDR. AIAs high-risk obligations will apply, which, while not identical, approach the overall compliance burden of a Class IIa medical device \textemdash~including NB involvement. Consequently, while Class I design flexibility remain, its financial and temporal advantages diminish.

\textbf{Architectural modularisation:} Another approach is architectural separation. Instead of qualifying the entire chatbot as a medical device, only the medically relevant functionality is brought within MDR scope. For example, a general well-being chatbot might invoke an MDR-certified PROM-collection component through an API only when clinically necessary. This can confine MDR obligations to the medical module while allowing the remaining system to be developed outside medical device regulation. Such separation can be strategically beneficial because MDR classification is based on the highest-risk function within an integrated system~\cite{MDR_AnnVIII}. For instance, if the health Q\&A falls under Class I, but the nudging functionality qualifies as Class IIb, the entire system is classified as Class IIb.

Modularisation may increase flexibility in design, iteration, and deployment for non-medical features. However, it is not a guaranteed regulatory loophole. Regulators may assess the system as a whole if components are tightly integrated or marketed as a unified clinical solution. Artificial separation that does not reflect actual use may therefore not withstand regulatory scrutiny. 

\section{Conclusion}
This paper has shown how seemingly small conversational design decisions in health chatbots, such as role, actionability, and proactivity, can decisively shape their regulatory qualification and risk classification. By treating MDR, GDPR, and the AI Act as first-order design constraints rather than post-hoc hurdles, our analysis illustrates how human-centred and regulatory-centred design can be aligned from the outset. Doing so makes it more likely that promising prototypes will not only perform well in research studies but can also be translated into safe and lawful clinical tools.

\begin{acks}
This work was funded by the European Union as part of the Horizon Europe research initiative RES-Q+ (grant number 101057603). Views and opinions expressed are, however, those of the author(s) only and do not necessarily reflect those of the European Union or the Health and Digital Executive Agency. Neither the European Union nor the granting authority can be held responsible for them. Generative AI was used for grammar and language editing.
\end{acks}

\bibliographystyle{ACM-Reference-Format}
\bibliography{sample-base}

\appendix

\end{document}